\documentclass[pra, twocolumn, floatfix, superscriptaddress, longbibliography, preprintnumbers]{revtex4-1}

\usepackage{amsmath, amssymb, bbm, bbold}
\usepackage{graphicx}
\usepackage{color}

\newcommand{\ket}[1]{{| #1 \rangle}}
\newcommand{\bra}[1]{{\langle #1 |}}

\newcommand{\Tr}{{\rm Tr}}
\newcommand{\sinc}{{\rm sinc}}

\begin{document}

\title{Jumptime unraveling of Markovian open quantum systems}

\author{Clemens Gneiting}
\email{clemens.gneiting@riken.jp}
\affiliation{Theoretical Quantum Physics Laboratory, RIKEN Cluster for Pioneering Research, Wako, Saitama 351-0198, Japan}
\affiliation{RIKEN Center for Quantum Computing, Wako, Saitama 351-0198, Japan}
\author{A.V. Rozhkov}
\affiliation{Institute for Theoretical and Applied Electrodynamics, Russian Academy of Sciences, Moscow, 125412 Russia}
\author{Franco Nori}
\affiliation{Theoretical Quantum Physics Laboratory, RIKEN Cluster for Pioneering Research, Wako, Saitama 351-0198, Japan}
\affiliation{RIKEN Center for Quantum Computing, Wako, Saitama 351-0198, Japan}
\affiliation{Department of Physics, University of Michigan, Ann Arbor, Michigan 48109-1040, USA}

\date{\today}

\begin{abstract}
We introduce jumptime unraveling as a distinct description of open quantum systems. As our starting point, we consider quantum jump trajectories, which emerge, physically, from continuous quantum measurements, or, formally, from the unraveling of Markovian quantum master equations. If the stochastically evolving quantum trajectories are ensemble-averaged at specific times, the resulting quantum states are solutions to the associated quantum master equation. We demonstrate that quantum trajectories can also be ensemble-averaged at specific jump counts. The resulting jumptime-averaged quantum states are then solutions to a discrete, deterministic evolution equation, with time replaced by the jump count. This jumptime evolution represents a trace-preserving quantum dynamical map if and only if the associated open system does not exhibit dark states. In the presence of dark states, on the other hand, jumptime-averaged states may decay into the dark states and the jumptime evolution may eventually terminate. Jumptime-averaged quantum states and the associated jumptime evolution are operationally accessible in continuous measurement schemes, when quantum jumps are detected and used to trigger the readout measurements. We illustrate the jumptime evolution with the examples of a two-level system undergoing relaxation or dephasing, a damped harmonic oscillator, and a free particle exposed to collisional decoherence.
\end{abstract}

\preprint{\textsf{published in Phys.~Rev.~A~{\bf 104}, 062212 (2021)}}

\maketitle

\section{Introduction}

An isolated, unobserved quantum system follows Schr\"odinger dynamics and thus describes a smooth, deterministic evolution in state space, much in the spirit of classical field theories. In order to retrieve information about the system, however, we must measure it. We can do so at chosen, isolated times, forcing the quantum state into instantaneous, abrupt changes to comply with specific measurement outcomes. These probabilistic changes, formalized in Born's rule, lie at the core of quantum mechanics, recasting it as a statistically predictive theory.

If a quantum system is {\it continuously} monitored, the measurement apparatus delivers an ongoing measurement record, now informing us about the stochastic evolution of the system, its {\it quantum trajectory}. This record/trajectory describes intervals of continuous, deterministic evolution, interrupted by sudden changes at random times, {\it quantum jumps}, such as, e.g., in continuous photon counting measurements. Depending on the specifics of the measurement process, part or even all of the probabilistic nature of the continuous measurement is then absorbed in the stochastically occurring jump times.

The theoretical foundation of quantum trajectories marks a milestone in the refinement of quantum theory \cite{Gisin1984quantum, Diosi1986stochastic, Belavkin1990stochastic, Carmichael1993open, Gardiner1992wave, Dalibard1992wave, Gisin1992quantum, Plenio1998quantum}, with numerous conceptual and practical implications (e.g., \cite{Mensky1993continuous, Braginsky1995quantum, Strunz1999open, Brun2000continuous, Doherty2000quantum, Wiseman2001complete, Piilo2008nonMarkovian, Busse2009pointer, Garrahan2010thermodynamics, Hofer2013time, Gammelmark2013past, Brandes2016feedback, Zhang2017quantum, Ashida2018thermalization, Manzano2018harnessing, Vincentini2019optimal, Smirne2020rate, Macri2021revealing}). For instance, continuous measurements are central for quantum feedback control \cite{Wiseman2009quantum, Jacobs2014quantum}. In experiment, quantum jumps and the associated individual quantum trajectories have been successfully traced on various platforms \cite{Nagourney1986shelved, Sauter1986observation, Bergquist1986obesrvation, Peil1999observing, Gustavsson2006counting, Fujisawa2006bidirectional, Gleyzes2007quantum, Kubanek2009photon, Neumann2010single, Vijay2011observation, Sayrin2011real, Pla2013high, Minev2019catch, Kurzmann2019optical}.

A deep conceptual relationship exists between quantum trajectories and Markovian open quantum systems, which can be described by Markovian quantum master equations: Any Markovian quantum master equation can be {\it unraveled} in terms of quantum trajectories, associated with a continuous measurement process, such that the ensemble average over all trajectories recovers the solution of the master equation. {\it Vice versa}, if measurement records are discarded in a continuous measurement, we must ensemble-average over all quantum trajectories, where the ensemble-averaged state then follows a quantum master equation.

\begin{figure}[htb]
\includegraphics[width=0.99\columnwidth]{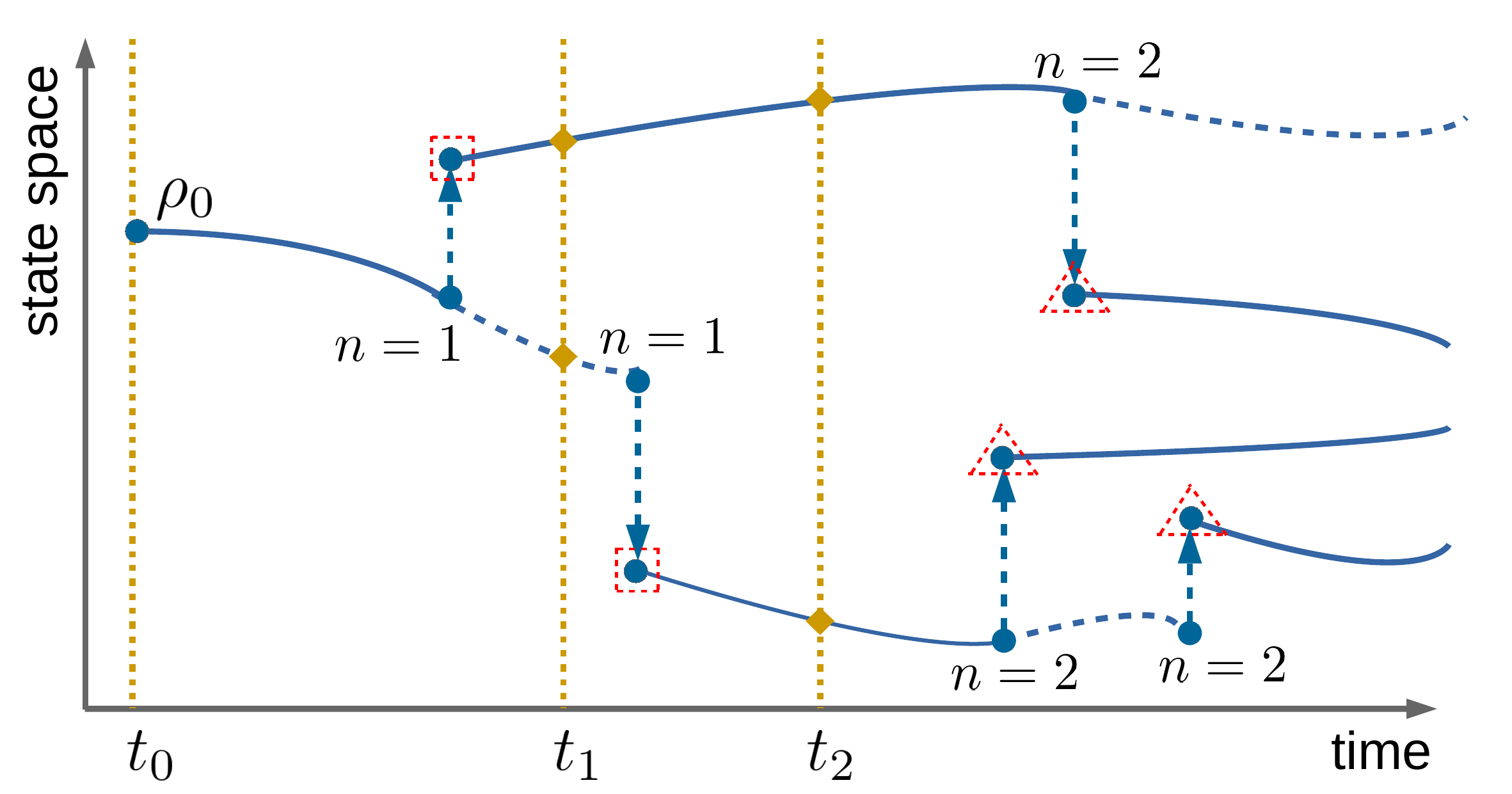}
\caption{\label{Fig:jump_unraveling} Quantum jump unraveling of Markovian open quantum systems. Quantum trajectories undergo piecewise deterministic time evolutions (blue solid and dashed lines, where the latter describe alternative trajectories), interrupted by stochastically occurring quantum jumps (blue dashed arrows). In walltime averaging, quantum trajectories are ensemble-averaged at fixed times $t_i$ (beige dotted lines), where different trajectories have in general accumulated different numbers of quantum jump events. In {\it jumptime averaging}, on the other hand, quantum trajectories are ensemble-averaged at fixed jump counts $n$, where different quantum trajectories in general arrive at $n$ quantum jumps at different times. The red dashed squares (triangles) refer to the fixed jump count $n=1$ ($n=2$), and these are averaged separately.}
\end{figure}

In standard {\it walltime} averaging of quantum jump trajectories, the latter are ensemble-averaged at fixed times, where different trajectories have in general accumulated different numbers of quantum jump events, cf.~Fig~\ref{Fig:jump_unraveling}. Here, we demonstrate that quantum trajectories can also be consistently ensemble-averaged by bundling them at fixed counts of jump events, i.e., at the {\it jump times}, see Fig.~\ref{Fig:jump_unraveling}. As we show, the such defined quantum states follow a discrete, deterministic evolution equation, retaining the resolution into jump events while stripping off the stochasticity of the jump occurrence. In this sense, this {\it jumptime unraveling} represents a distinct way of analyzing open quantum systems, or, for that matter, continuous quantum measurements.

We would like to stress that we use the term ``jump time'' in the interpretation of ``time point of occurrence of a quantum jump'', whereas the same term can also be used in the sense of ``duration of a quantum jump'', see, e.g., \cite{Schulman2002jump}. Note, however, that, in the context of quantum trajectories, quantum jumps necessarily happen instantaneously, and hence there is no risk of confusing these two interpretations.

This article is structured as follows: In Section II, we briefly recapitulate quantum jump trajectories and how they are related to quantum master equations through walltime averaging. In Section III, we then introduce the jumptime averaging of quantum trajectories and elaborate the corresponding readout protocol in monitoring implementations. The discrete, deterministic jumptime evolution equation, which connects subsequent jumptime-averaged quantum states, is derived in Section IV. Section V then elaborates a theorem that relates the trace preservation of the jumptime evolution equation to the presence or absence of dark states, while the waiting time distribution between subsequent jumptime-averaged quantum states is presented in Section VI. In Section VII, we discuss several basic examples which serve to demonstrate some of the characteristic consequences of the jumptime evolution, followed by our conclusions in Section VIII.

\section{Quantum trajectories}

Let us consider an open quantum system characterized by a Hamiltonian $\hat{H}$ and a collection of jump/Lindblad operators $\{ \hat{L}_j| j \in \mathcal{I} \}$. Its continuous-time (or {\it walltime}) evolution is then governed by a Markovian (Gorini-Kossakowski-Sudarshan-Lindblad) quantum master equation \cite{Gorini1976completely, Lindblad1976generators},
\begin{align} \label{Eq:Lindblad_master_equation}
\partial_t \rho_t = -\frac{i}{\hbar} [\hat{H}, \rho_t] + \gamma \sum_{j \in \mathcal{I}} \Big( \hat{L}_j \rho_t \hat{L}_j^\dagger - \frac{1}{2} \{ \hat{L}_j^\dagger \hat{L}_j, \rho_t \} \Big) ,
\end{align}
where $\{ \hat{A},\hat{B} \} = \hat{A} \hat{B} + \hat{B} \hat{A} $. The Lindblad operators $\hat{L}_j$ account for incoherent contributions to the dynamics, e.g., induced by an environment. In the context of continuous measurements, the jump operators are specified by the nature of the measurement process, which can be characterized by the positive operator-valued measure (POVM) $\hat{F}_j = \gamma dt \hat{L}_j^\dagger \hat{L}_j, \, j \in \mathcal{I}$, and $\hat{\overline{F}} = \mathbb{1} - \gamma dt \sum_{j \in \mathcal{I}} \hat{L}_j^\dagger \hat{L}_j$, where $\hat{\overline{F}} + \sum_{j \in \mathcal{I}} \hat{F}_j = \mathbb{1}$. Accordingly, we may say in the former case that ``the environment measures/monitors the system''. Due to the presence of the ``null outcome'' $\hat{\overline{F}}$, such measurements are sometimes called {\it weak} continuous measurements. Note that $\gamma>0$ has the dimension of a rate.

Quantum jump unraveling interprets the solution $\rho_t$ of Eq.~(\ref{Eq:Lindblad_master_equation}) as emerging from the ensemble average over stochastically evolving quantum trajectories. The unraveled solution can then be written as \cite{Carmichael1993open}
\begin{align} \label{Eq:jump_expansion}
\rho_t = \sum_{n=0}^{\infty} \int_{0}^{t} \!\! dt_n \int_{0}^{t_n} \!\! dt_{n-1} \dots \int_{0}^{t_2} \!\! dt_{1} \!\!\!\! \sum_{j_1, \dots j_n \in \mathcal{I}} \!\! \rho^t_{j_n \dots j_1}(\{t_i\}) ,
\end{align}
where the (unnormalized) quantum trajectories
\begin{align} \label{Eq:quantum_trajectory}
\rho^t_{j_n \dots j_1}(\{t_i\}) = \mathcal{U}_{t-t_n} \mathcal{J}_{j_n} \mathcal{U}_{t_n-t_{n-1}} \mathcal{J}_{j_{n-1}} \dots \mathcal{J}_{j_{1}} \mathcal{U}_{t_1} \rho_0
\end{align}
describe the non-unitary time evolution of an (in general mixed) initial state $\rho_0$, conditioned on $n$ quantum jumps of type $j_1, j_2, \dots, j_n$ occurring at times $t_1, t_2, \dots, t_n$. Here, we define the deterministic time evolution between jumps $\mathcal{U}_t \rho \equiv e^{-\frac{i}{\hbar} \hat{H}_{\rm eff} t} \rho e^{\frac{i}{\hbar} \hat{H}_{\rm eff}^\dagger t}$ (corresponding to the null outcome $\hat{\overline{F}}$), the jump effects $\mathcal{J}_j \rho \equiv \gamma \hat{L}_j \rho \hat{L}_j^\dagger$ (corresponding to the measurement outcomes $\hat{F}_j$), and the (non-Hermitian) effective Hamiltonian
\begin{align} \label{Eq:effective_Hamiltonian}
\hat{H}_{\rm eff} = \hat{H} - i \hbar \frac{\gamma}{2} \sum_{j \in \mathcal{I}} \hat{L}^{\dagger}_j \hat{L}_j .
\end{align}
Let us clarify that the $n=0$ term in the sum (\ref{Eq:jump_expansion}) describes the evolution conditioned on no quantum jumps occurring, $\mathcal{U}_{t} \rho_0$. Moreover, we stress that the jump events $\mathcal{J}_j$ in individual quantum trajectories (\ref{Eq:quantum_trajectory}) specify single jump types $j \in \mathcal{I}$, while the ensemble average (\ref{Eq:jump_expansion}) recovers the sum over all jump types featured in (\ref{Eq:Lindblad_master_equation}). In continuous monitoring implementations, we can think of the quantum jumps as being recorded as ``clicks'' in the detector.

The quantum master equation (\ref{Eq:Lindblad_master_equation}) is known (see, e.g., \cite{Wiseman2009quantum}) to be invariant under the transformation
\begin{subequations} \label{Eq:Lindblad_invariance}
\begin{align}
\hat{L}_j &\rightarrow \hat{L}_j' = \hat{L}_j + \alpha_j , \\
\hat{H} &\rightarrow \hat{H}' = \hat{H} - i \hbar \frac{\gamma}{2} \sum_{j \in \mathcal{I}} \left( \alpha_j^* \hat{L}_j - \alpha_j \hat{L}_j^\dagger \right) ,
\end{align}
\end{subequations}
where $\alpha_j \in \mathbb{C}.$ Correspondingly, we obtain for each choice of the displacements $\alpha_j$ a different unraveling (\ref{Eq:jump_expansion}) of the quantum master equation (\ref{Eq:Lindblad_master_equation}), each associated with a different continuous measurement that opens the system in a different way. This transformation underlies, e.g., homodyne measurements. With increasing displacements $\alpha_j$, the frequency of quantum jumps increases, while the effect of individual quantum jumps decreases, so that it can be useful (or experimentally required) to subsume them by a time-continuous diffusion process. Such effective diffusive unravelings (which can always be reduced with arbitrary accuracy to punctuated quantum jumps \cite{Wiseman2001complete}) are excluded here. Ensemble averages over classical noise/disorder realizations \cite{Vega2017dynamics, Gneiting2020disorder}, which are not related to continuous quantum measurements, are excluded, too.

\section{Jumptime averaging}

In the unraveling (\ref{Eq:jump_expansion}) of the walltime evolution equation (\ref{Eq:Lindblad_master_equation}), quantum trajectories are averaged over at specific (wall)times $t$. In general, $\rho_t$ then contains contributions from any jump order $n$. {\it Jumptime unraveling} is based on the insight that quantum trajectories can alternatively be bundled together at a specified jump order $n$, i.e., by averaging over quantum trajectories immediately after they have completed $n$ jump events (irrespective of their jump types). Importantly, different trajectories in general arrive at the $n$th jump at different times. Time $t$ is then replaced by the jump order $n$, $\rho_t \rightarrow \rho_n$, and
\begin{align} \label{Eq:jump-time_unraveling}
\rho_n = \int_{0}^{\infty} \!\! dt_n \int_{0}^{t_n} \!\! dt_{n-1} \dots \int_{0}^{t_2} \!\! dt_{1} \!\!\!\! \sum_{j_1, \dots j_n \in \mathcal{I}} \rho^n_{j_n \dots j_1}(\{t_i\}) ,
\end{align}
with the modified quantum trajectories
\begin{align} \label{Eq:jump-time_conditioned_density_operators}
\rho^n_{j_n \dots j_1}(\{t_i\}) = \mathcal{J}_{j_n} \mathcal{U}_{t_n-t_{n-1}} \mathcal{J}_{j_{n-1}} \dots \mathcal{J}_{j_{1}} \mathcal{U}_{t_1} \rho_0 .
\end{align}
Note that the upper limit of the integral over the final jump time $t_n$ in (\ref{Eq:jump-time_unraveling}) is shifted to infinity, as the occurrence of the last jump is now unconstrained in time. Moreover, the quantum trajectories (\ref{Eq:jump-time_conditioned_density_operators}) terminate immediately after the indicated number of jumps is reached.

By construction, the jumptime-averaged state $\rho_n$ describes a well-defined, positive semi-definite quantum state, which is normalized as long as every quantum trajectory reaches $n$ jumps. Moreover, it is operationally accessible in continuous monitoring schemes, where quantum trajectories are traced and quantum jumps detected. To see this, we relate the jumptime-averaged state $\rho_n$ to a readout protocol, which operationally explains how $\rho_n$ describes the statistics of readout measurements. Note that all quantum states refer, implicitly or explicitly, to a readout protocol. However, while commonly the time $t$ of the readout measurement is fixed, i.e., a quantum state $\rho_t$ describes the statistics of, actual or hypothetical, repeated measurements at the time $t$, $\rho_n$ is based on a readout protocol that connects different readout times.

\begin{figure}[htb]
	\includegraphics[width=0.99\columnwidth]{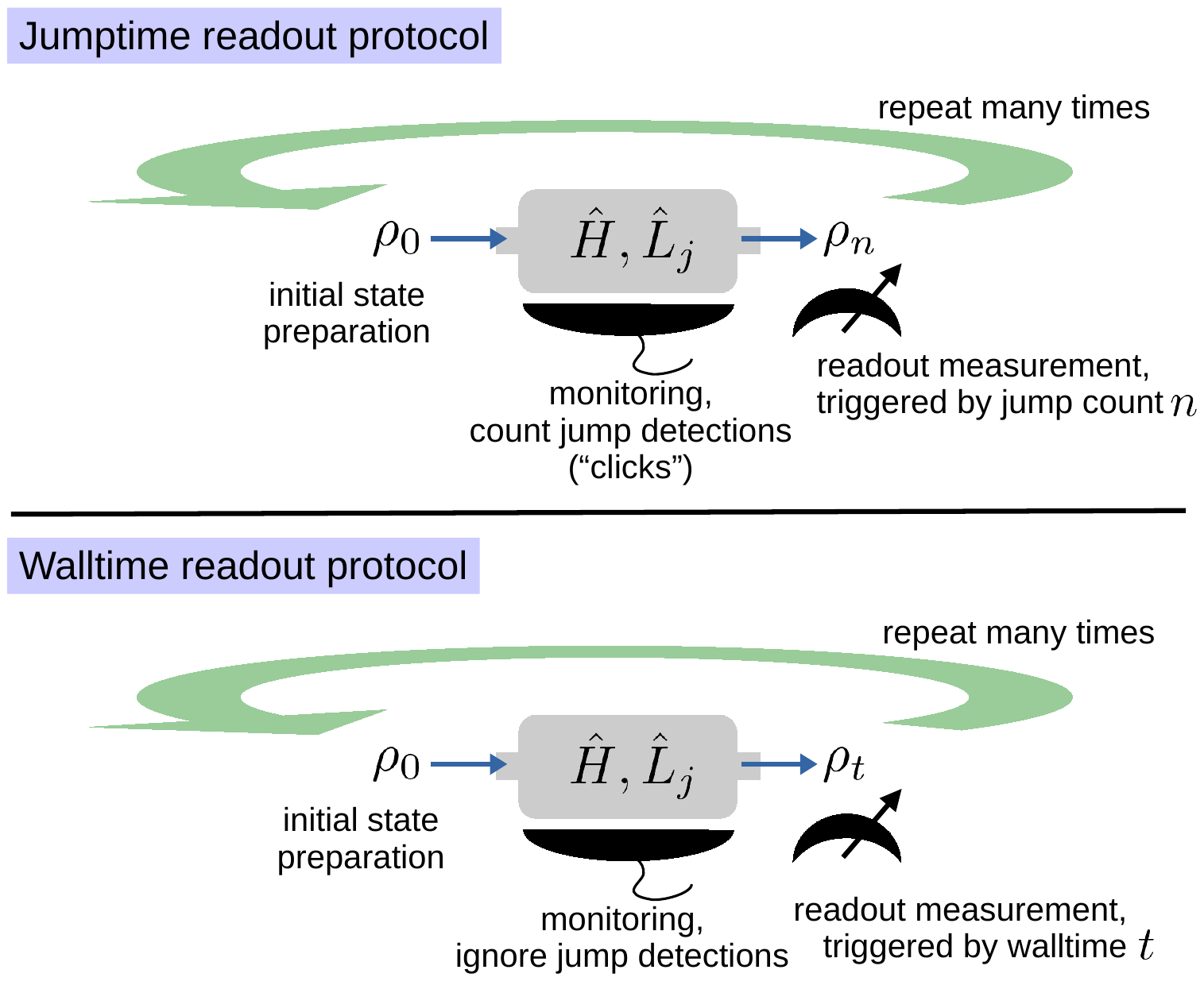}
	\caption{\label{Fig:readout_protocol} Readout protocols corresponding to the jumptime and the walltime averaging under monitoring, respectively. If the readout measurement is triggered by a preset jump count $n$ (upper panel), then the quantum state $\rho_n$ that captures the statistics of the readout measurement is a solution to the jumptime evolution equation (\ref{Eq:jumptime_evolution_equation}). On the other hand, if the readout measurement is triggered by a preset walltime $t$ (lower panel), then the quantum state $\rho_t$ that captures the statistics of the readout measurement is a solution to the Lindblad equation (\ref{Eq:Lindblad_master_equation}). The readout measurement can refer to an arbitrary observable $\hat{O} = \hat{O}^\dagger$, whose statistics is then described by $\langle \hat{O} \rangle_n = \Tr [\rho_n \hat{O}]$ in the case of jumptime averaging, and $\langle \hat{O} \rangle_t = \Tr [\rho_t \hat{O}]$ in the case of walltime averaging, respectively. More generally, the readout measurement can stand for any positive operator-valued measure. If one conducts an informationally complete set of measurements, then the full quantum state can be reconstructed, i.e., $\rho_n$ in the case of jumptime and $\rho_t$ in the case of walltime averaging, respectively.}
\end{figure}

The {\it readout protocol} for $\rho_n$, which involves knowledge of the jump occurrences, is as follows: After preparing the initial state $\rho_0$, continuously measure the system and count the quantum jumps (``clicks'' in the monitoring instrument) until the desired jump index $n$ is reached (the readout condition), then perform a concluding (strong) readout measurement. After repeating this many times, the statistics of the readout measurement is described by $\rho_n$, cf.~(\ref{Eq:jump-time_unraveling}). In contrast, if the final readout measurement is, regardless of the jump count, performed at a fixed time $t$ for each run (i.e., the jump detections are effectively discarded under this readout condition), then the statistics of the readout measurement is described by $\rho_t$, cf.~(\ref{Eq:jump_expansion}). The readout protocols underlying the jumptime and the walltime averaging, respectively, are illustrated in Fig.~\ref{Fig:readout_protocol}.

Let us note that quantum jump detections have previously been proposed as a trigger in quantum feedback protocols (see, e.g., \cite{Carvalho2007stabilizing}, and \cite{Wiseman2009quantum, Jacobs2014quantum} and reference therein); however, always in combination with the readout triggered by walltimes, i.e., walltime averaging. As we show in the remainder, jumptime averaging, i.e., the readout triggered by jump counts, gives rise to significantly distinct dynamical characteristics.

\section{Jumptime evolution}

The jumptime-averaged state $\rho_n$ can be determined without resorting to individual quantum trajectories. To see this, we rewrite (\ref{Eq:jump-time_unraveling}) by switching to waiting times $\tau_n = t_n - t_{n-1}$ between jumps, which yields
\begin{align}
\rho_n &= \int_{0}^{\infty} \!\! d\tau_n \int_{0}^{\infty} \!\! d\tau_{n-1} \dots \int_{0}^{\infty} \!\! d\tau_{1} \!\!\!\! \sum_{j_1, \dots j_n \in \mathcal{I}} \rho^n_{j_n \dots j_1}(\{\tau_i\}) ,
\end{align}
where all waiting times decouple and extend to infinity, as the time ordering required in the walltime unraveling (\ref{Eq:jump_expansion}) is here ineffective. It is now straightforward to extract the recursive relation
\begin{align} \label{Eq:jumptime_evolution_equation}
\rho_{n+1} = \int_{0}^{\infty} \!\! \gamma d\tau \sum_{j \in \mathcal{I}} \hat{L}_j e^{-\frac{i}{\hbar} \hat{H}_{\rm eff} \tau} \rho_n e^{\frac{i}{\hbar} \hat{H}_{\rm eff}^\dagger \tau} \hat{L}_j^\dagger ,
\end{align}
where we have returned to standard operator notation. Repetitive application of (\ref{Eq:jumptime_evolution_equation}) allows us now to determine the jumptime-evolved state $\rho_n$ for any $n \ge 1$, starting from an arbitrary initial state $\rho_0$. We emphasize that (\ref{Eq:jumptime_evolution_equation}), which describes a completely positive quantum map in Kraus form, is derived without approximations (i.e., it is exact), and is applicable to general open quantum systems.

The jumptime evolution equation (\ref{Eq:jumptime_evolution_equation}) is our main result. Based on a distinct, operationally meaningful protocol to bundle the stochastically evolving quantum trajectories, we obtain a discrete, deterministic evolution, preserving the quantum jumps while absorbing their stochasticity in the jump order. The jumptime evolution (\ref{Eq:jumptime_evolution_equation}) replaces the walltime evolution (\ref{Eq:Lindblad_master_equation}) for jumptime-averaged quantum states (\ref{Eq:jump-time_unraveling}), and the latter unravel the jumptime evolution equation (\ref{Eq:jumptime_evolution_equation}) in the same way as the walltime-averaged states (\ref{Eq:jump_expansion}) unravel the Lindblad evolution equation (\ref{Eq:Lindblad_master_equation}). As such, this jumptime unraveling provides a novel window into open quantum systems.

We emphasize that experimental tests of the jumptime evolution (\ref{Eq:jumptime_evolution_equation}) through monitoring, following the above readout protocol, do not involve conditioning (apart from the readout condition, which is always required) or postselection, as all quantum trajectories contribute to the jumptime-averaged state (\ref{Eq:jump-time_unraveling}). In contrast, the repeated readout of specific quantum trajectories, which are conditioned on both the sequence of jump types and their times of occurrence, requires postselection, which rapidly becomes prohibitive with an increasing number of quantum jumps. The same holds for the conditioning on subensembles of quantum trajectories, or, generally, whenever the readout condition is complemented by additional conditioning, where we cannot control all conditions simultaneously.
	
Moreover, we clarify that the time integral to infinity in (\ref{Eq:jumptime_evolution_equation}), which reflects the stochastic occurrence of the quantum jumps, does not describe the need for infinite observation times. While it can, in principle, take arbitrarily long for the subsequent quantum jump to occur, this will always happen at finite times (unless quantum jumps have entirely ceased, as discussed in the following section). Correspondingly, an arbitrary jump count $n$ is always reached in finite time. In experimental realizations, one may prefer, or be required, to introduce a maximum waiting time $T$ in between jumps (e.g., to limit the overall maximum duration of individual runs), where a run is discarded if the waiting time after an intermediate jump exceeds this threshold. If the cutoff time $T$ can be chosen sufficiently large to lie in the exponential tails of all involved waiting time distributions [cf.~(\ref{Eq:waiting_time_distribution})], the discarded fraction of quantum trajectories is negligible. Otherwise, a modified (not norm-preserving) jumptime evolution equation with the upper integration limit replaced by $T$ would apply.

It is also worth mentioning that, similar to the Lindblad evolution (\ref{Eq:Lindblad_master_equation}), finite samples of quantum trajectories can be used to approximate the jumptime-averaged states (\ref{Eq:jump-time_unraveling}), and thus to efficiently simulate the jumptime evolution (\ref{Eq:jumptime_evolution_equation}) by propagating pure states.

Finally, let us point out that, in the generic case, the jumptime evolution (\ref{Eq:jumptime_evolution_equation}) lifts the degeneracy (\ref{Eq:Lindblad_invariance}) of the Lindblad equation (\ref{Eq:Lindblad_master_equation}). This ability to discriminate can be attributed to the fact that the jumptime averaging keeps track of the quantum jumps, while the latter are discarded (and hence ``washed out'') in the walltime averaging. A family of examples, where the lifting of the degeneracy (\ref{Eq:Lindblad_invariance}) by the jumptime evolution becomes apparent, will be given at the end of Section \ref{Sec:role_of_dark_states}. Note, however, that the jumptime evolution equation (\ref{Eq:jumptime_evolution_equation}) may be subject to other degeneracies. In the (diffusive) limit of $\alpha_j \rightarrow \infty$ (for at least one $j$), where quantum jumps occur in arbitrarily close succession, jumptime and walltime averaging become effectively indistinguishable, and the jumptime evolution reduces to the walltime evolution.

\section{\label{Sec:role_of_dark_states}Role of dark states}

In general, the jumptime evolution (\ref{Eq:jumptime_evolution_equation}) is non-trace-increasing, i.e., $\Tr[\rho_{n+1}] \leq \Tr[\rho_n] \le \Tr[\rho_0] = 1$ for any $\rho_0$, where the norm of $\rho_{n+1}$ is reduced by the fraction of quantum trajectories whose chain of quantum jumps terminates after $n$ jumps. We now prove that, for systems with finite-dimensional state space, the jumptime evolution (\ref{Eq:jumptime_evolution_equation}) describes a {\it trace-preserving} quantum channel, i.e., any quantum trajectory exhibits infinitely many quantum jumps regardless of the initial state $\rho_0$, if and only if the master equation (\ref{Eq:Lindblad_master_equation}) does not exhibit dark states.

A pure state $|\psi_{\rm D} \rangle$ constitutes a dark state, if it lies in the kernel of all jump operators,
\begin{subequations}
\begin{align}
\hat{L}_j |\psi_{\rm D} \rangle = 0 \,\,\,\, \forall j \in \mathcal{I} ,
\end{align}
and, simultaneously, the projector $|\psi_{\rm D} \rangle \langle \psi_{\rm D}|$ commutes with the Hamiltonian:
\begin{align}
[\hat{H}, |\psi_{\rm D} \rangle \langle \psi_{\rm D}|] = 0 .
\end{align}
\end{subequations}
Dark states thus describe pure steady states of the master equation (\ref{Eq:Lindblad_master_equation}) that, once reached, are characterized by quantum trajectories that do not exhibit quantum jumps. It follows immediately that the jumptime evolution ends at a dark state: $\rho_{n+1} = 0$ if $\rho_n = |\psi_{\rm D} \rangle \langle \psi_{\rm D}|$; i.e., dark states define the kernel of the jumptime evolution (\ref{Eq:jumptime_evolution_equation}).

A quantum/Kraus map \cite{Kraus1971general} $\rho \rightarrow \rho' = \varepsilon(\rho) = \sum_{\mu} E_{\mu} \rho E_{\mu}^{\dagger}$ is completely positive and trace-preserving, if the operators $E_{\mu}$ satisfy the completeness relation $\sum_{\mu} E_{\mu}^{\dagger} E_{\mu} = \mathbb{1}$. This leaves us to show that
\begin{align} \label{Eq:Kraus_identity}
S \equiv \int_{0}^{\infty} \!\! \gamma d\tau \, e^{\frac{i}{\hbar} \hat{H}_{\rm eff}^\dagger \tau} \hat{V} e^{-\frac{i}{\hbar} \hat{H}_{\rm eff} \tau} = \mathbb{1}
\end{align}
in the absence of dark states, where we have introduced the effective potential $\hat{V} = \hat{V}^{\dagger} = \sum_{j \in \mathcal{I}} \hat{L}_j^\dagger \hat{L}_j$. Note that $\hat{V}$ is non-negative definite.

It is instructive to first consider the case $\hat{H}=0$. If we spectrally decompose $\hat{V} = \sum_m v_m |m \rangle \langle m|$, where $v_m \ge 0$, we can write $S = \sum_m v_m \left\{ \int_{0}^{\infty} \gamma d\tau \, e^{-\gamma v_m \tau} \right\} |m\rangle \langle m|$. If $v_m>0$ $\forall m$ (no dark states), the integrals yield $v_m^{-1}$ and we recover the identity operator. However, if there exists an $m$ such that $v_m=0$ (dark state), the associated term must be excluded from the sum and the identity operator cannot be recovered, implying that the quantum map is not trace-preserving.

In the general case, $\hat{H} \neq 0$, we start by noting that $\gamma \hat{V} = -\frac{i}{\hbar} (\hat{H}_{\rm eff}^{\dagger} - \hat{H}_{\rm eff})$, which allows us to rewrite the integrand in (\ref{Eq:Kraus_identity}) as a total derivative,
\begin{align}
-S = \int_{0}^{\infty} \!\!\!\! d\tau \frac{d}{d\tau} \left[ e^{\frac{i}{\hbar} \hat{H}_{\rm eff}^\dagger \tau} e^{-\frac{i}{\hbar} \hat{H}_{\rm eff} \tau} \right] = e^{\frac{i}{\hbar} \hat{H}_{\rm eff}^\dagger \tau} e^{-\frac{i}{\hbar} \hat{H}_{\rm eff} \tau} \Big|_0^{\infty} \nonumber ,
\end{align}
where we formally performed the integral. The integral exists and Eq.~(\ref{Eq:Kraus_identity}) is validated, if and only if
\begin{align} \label{Eq:vanishing_limit}
\lim\limits_{\tau \rightarrow \infty} e^{\frac{i}{\hbar} \hat{H}_{\rm eff}^\dagger \tau} e^{-\frac{i}{\hbar} \hat{H}_{\rm eff} \tau} = 0 .
\end{align}

We proceed with proving that Eq.~(\ref{Eq:vanishing_limit}) holds if and only if there are no dark states. First, we realize that there always exists a (finite-dimensional) canonical basis $|\lambda \rangle$, in which $\hat{H}_{\rm eff}^{\dagger}$ takes Jordan-normal form. Recall that every eigenvalue comes with at least one ordinary eigenstate. Moreover, all eigenvalues of $\hat{H}_{\rm eff}^{\dagger}$ have non-negative imaginary parts. Indeed, if $|\lambda \rangle$ is an ordinary eigenstate of $\hat{H}_{\rm eff}^{\dagger}$, $\hat{H}_{\rm eff}^{\dagger} |\lambda \rangle = z_{\lambda} |\lambda \rangle$, we obtain $\langle \lambda| \hat{H}_{\rm eff}^{\dagger} |\lambda \rangle = z_{\lambda} = \langle \lambda| \hat{H} |\lambda \rangle + i \hbar \frac{\gamma}{2} \langle \lambda| \hat{V} |\lambda \rangle$. Since both $\hat{H}$ and $\hat{V}$ are Hermitian and thus real on the diagonal, and since $\hat{V}$ is in addition positive semi-definite, then ${\rm Im} \, z_{\lambda} = \hbar \frac{\gamma}{2} \langle \lambda| \hat{V} |\lambda \rangle \ge 0$ follows immediately.

We now show that an ordinary eigenstate $|\lambda \rangle$ of $\hat{H}_{\rm eff}^{\dagger}$ is a dark state if and only if ${\rm Im} \, z_{\lambda} = 0$. One direction is trivial: If $|\lambda \rangle$ is a dark state, then by definition $\langle \lambda| \hat{V} |\lambda \rangle = 0$, and hence ${\rm Im} \, z_{\lambda} = 0$. On the other hand, if $\langle \lambda| \hat{V} |\lambda \rangle = \sum_{j \in \mathcal{I}} \langle \lambda| \hat{L}_j^\dagger \hat{L}_j |\lambda \rangle = 0$, we infer $\hat{L}_j |\lambda \rangle = 0$ $\forall j$. Since $|\lambda \rangle$ is an ordinary eigenstate of $\hat{H}_{\rm eff}^{\dagger}$, we also infer $\hat{H}_{\rm eff}^{\dagger} |\lambda \rangle = \hat{H} |\lambda \rangle = z_{\lambda} |\lambda \rangle$, that is, $|\lambda \rangle$ is also an eigenstate of $\hat{H}$ and hence a dark state.

Now let us assume that a state $|\mu \rangle$ is a dark state. Then $\hat{V} |\mu \rangle = 0$ and $\hat{H}_{\rm eff}^{\dagger} \ket{\mu} = \hat{H}_{\rm eff} \ket{\mu} = \hat{H} \ket{\mu} = \varepsilon \ket{\mu}$ with $\varepsilon \in \mathbb{R}$, i.e., $\ket{\mu}$ is an eigenstate of both $\hat{H}_{\rm eff}^{\dagger}$ and $\hat{H}_{\rm eff}$. Consequently,
\begin{align}
\lim\limits_{\tau \rightarrow \infty} \bra{\mu} e^{\frac{i}{\hbar} \hat{H}_{\rm eff}^\dagger \tau} e^{-\frac{i}{\hbar} \hat{H}_{\rm eff} \tau} \ket{\mu} = \bra{\mu} \mu \rangle = 1 ,
\end{align}
which contradicts (\ref{Eq:vanishing_limit}).

On the other hand, if there are no dark states, we can infer that all eigenvalues of $\hat{H}_{\rm eff}^{\dagger}$ have strictly positive imaginary parts. If $\hat{j}_{\lambda}$ denotes the Jordan block associated with the eigenvalue $z_{\lambda}$, we can write
\begin{align}
e^{i \hat{j}_{\lambda} \tau} = e^{i z_{\lambda} \tau} \hat{F}_{\lambda}(\tau) = e^{- {\rm Im} z_{\lambda} \tau} \left( e^{i {\rm Re} z_{\lambda} \tau} \hat{F}_{\lambda}(\tau) \right) ,
\end{align}
where the matrix function $\hat{F}_{\lambda}(\tau)$ scales at most polynomially in $\tau$ for large $\tau$. Thus, at large $\tau$, all exponents of Jordan blocks $e^{i \hat{j}_{\lambda} \tau}$ decay to zero. Consequently, $e^{\frac{i}{\hbar} \hat{H}_{\rm eff}^\dagger \tau}$ vanishes in the limit $\tau \rightarrow \infty$. The same holds for $e^{-\frac{i}{\hbar} \hat{H}_{\rm eff} \tau}$, which can be proven similarly. Therefore, Eq.~(\ref{Eq:vanishing_limit}) is fulfilled in the absence of dark states. This completes our proof. \hfill $\blacksquare$

The relation between dark states and trace preservation does not necessarily hold in infinite-dimensional state spaces. Counterexamples, where quantum jumps are excluded even in the absence of dark states, can be constructed, e.g., if the dissipation acts in a locally confined region of space while the wave packet escapes in opposite direction to infinity. Infinite-dimensional systems with block-diagonal effective Hamiltonians $\hat{H}_{\rm eff} = \sum_k \ket{k}\bra{k} \otimes \hat{H}_{\rm eff}(k)$, where all blocks $\hat{H}_{\rm eff}(k)$ are finite dimensional, are trace-preserving exactly if all blocks are free of dark states. This corollary of the above theorem covers, e.g., translation-invariant models and both infinite-dimensional instances discussed below.

In the presence of dark states, decaying jumptime states $\rho_n$ indicate convergence towards the dark states, i.e., the steady states of the corresponding walltime master equations. In particular, if there is a single dark state and $\rho_n = 0$, then the system state is the dark state. If there are multiple dark states, then $\rho_n = 0$ indicates that the system state is confined to a (decoherence-free) subspace spanned by the dark states. Note, however, that the existence of dark states does not necessarily imply a decaying jumptime evolution, which, in general, depends on the initial state.

Finally, we remark that open quantum systems which exhibit dark states provide simple examples where the jumptime evolution in general lifts the degeneracy (\ref{Eq:Lindblad_invariance}) of the Lindblad evolution: Any nonvanishing choice of $\alpha_j$ in (\ref{Eq:Lindblad_invariance}) will remove the dark states, and hence will necessarily result in a perpetual jumptime evolution, independent of the initial state. In contrast, in the presence of dark states, there always exist initial conditions which result in a terminating jumptime evolution.

\section{Waiting time distribution}

If the jumptime evolution (\ref{Eq:jumptime_evolution_equation}) is trace-preserving, the trace of the integrand in (\ref{Eq:jumptime_evolution_equation}) delivers the waiting time distribution $w_{n \rightarrow n+1}(\tau)$ between jump $n$ and jump $n+1$,
\begin{align} \label{Eq:waiting_time_distribution}
w_{n \rightarrow n+1}(\tau) = \gamma \Tr \Big[ e^{\frac{i}{\hbar} \hat{H}_{\rm eff}^\dagger \tau} \sum_{j \in \mathcal{I}} \hat{L}_j^\dagger \hat{L}_j e^{-\frac{i}{\hbar} \hat{H}_{\rm eff} \tau} \rho_n \Big] .
\end{align}
Indeed, $\Tr[\hat{L}_j e^{-\frac{i}{\hbar} \hat{H}_{\rm eff} \tau} \rho_n e^{\frac{i}{\hbar} \hat{H}_{\rm eff}^\dagger \tau} \hat{L}_j^\dagger] \ge 0$ $\forall j, \tau$, and $\int_{0}^{\infty} d\tau \, w_{n \rightarrow n+1}(\tau) = \Tr \rho_{n+1}=1$. The non-negativity follows from the general consideration $\Tr[\hat{A} \rho \hat{A}^\dagger] = \sum_n p_n \bra{n} \hat{A}^\dagger \hat{A} \ket{n}$, where $\rho = \sum_n p_n \ket{n}\bra{n}$ with $p_n \ge 0$, and $\hat{A}^\dagger \hat{A}$ is non-negative. Note that the waiting time distribution, which is exact and general, is not restricted to stationary states and may vary substantially between different jump orders, depending on whether $\rho_n$ resides in a long- or a short-lived state. For $n=0$, expression~(\ref{Eq:waiting_time_distribution}) describes the waiting time distribution between the preparation of the initial state and the first quantum jump.

A state-independent waiting time distribution, $w_{n \rightarrow n+1}(\tau) = w(\tau)$, is obtained when $\hat{V} = \sum_{j \in \mathcal{I}} \hat{L}_j^\dagger \hat{L}_j = \mathbb{1}$, i.e., when the effective potential acts state-independently. The waiting time distribution then simplifies to
\begin{align} \label{Eq:state-independent_waiting_time}
w(\tau) = \gamma e^{-\gamma \tau} ,
\end{align}
which can be related to other methods to analyze the counting statistics of quantum jumps \cite{Brandes2008waiting, Rudge2019counting}. Below we discuss two examples of this kind, a two-level system undergoing dephasing, and a free particle exposed to collisional decoherence.

\section{Examples}

We now demonstrate with a few basic examples some of the characteristics of the jumptime evolution.

\subsection{Two-level system undergoing relaxation}

Our first example is a two-level system exposed to relaxation, characterized by a single jump operator
\begin{align}
\hat{L} = \hat{\sigma}_- = \ket{0} \bra{1} ,
\end{align}
where $\ket{0}$ and $\ket{1}$ denote the ground and the excited state, respectively. A general Hamiltonian can be written as $\hat{H} = \vec{h} \cdot \vec{\sigma}$, with the Pauli operators $\{ \hat{\sigma}_x, \hat{\sigma}_y, \hat{\sigma}_z \}$ and the convention $\hat{\sigma}_z = \ket{1}\bra{1} - \ket{0}\bra{0}$. The effective Hamiltonian (\ref{Eq:effective_Hamiltonian}) then reads
\begin{align}
\hat{H}_{\rm eff} = -i \hbar \frac{\gamma}{4} \mathbb{1}_2 + \vec{h}_{\rm eff} \cdot \vec{\sigma} ,
\end{align}
with $\vec{h}_{\rm eff} = (h_x, h_y, h_z-i \hbar \frac{\gamma}{4})^T$. Clearly, the state $\ket0$ is a dark state of the system if and only if $\hat{H} = h_z \hat{\sigma}_z$. Indeed, if $\hat{H} = 0$, the evolution operator representing the deterministic dynamics between consecutive jumps is equal to
\begin{align}
e^{-\frac{i}{\hbar} \hat{H}_{\rm eff} \tau} = \ket{0}\bra{0} + e^{-\frac{\gamma}{2} \tau} \ket{1}\bra{1} ,
\end{align}
and we obtain for the jumptime evolution (\ref{Eq:jumptime_evolution_equation})
\begin{align}
\rho_{n+1} = \bra{1} \rho_n \ket{1} \ket{0} \bra{0} .
\end{align}
It follows immediately that $\rho_{1} = \bra{1} \rho_0 \ket{1} \ket{0} \bra{0}$ and $\rho_2 = 0$, i.e., the jumptime evolution ends after the first jump latest. The vanishing jumptime state indicates that the system has reached the dark state $\ket{0}$, which is also the steady state of the corresponding walltime evolution equation $\partial_t \rho_t = \gamma \big( \hat{\sigma}_- \rho_t \hat{\sigma}_+ - \frac{1}{2} \{ \hat{\sigma}_+ \hat{\sigma}_-, \rho_t \} \big)$; the latter is solved by $\bra{0} \rho_t \ket{1} = \bra{0} \rho_0 \ket{1} \exp(-\gamma t/2)$ and $\bra{0} \rho_t \ket{0} = 1- \bra{1} \rho_0 \ket{1} \exp(-\gamma t)$.

On the other hand, if $\hat{H} = h_x \hat{\sigma}_x$ ($h_x \neq 0$), then $[\hat{H}, \ket{0} \bra{0}] \neq 0$ and the dark state is removed. Let us examine the special case $h_x = \hbar \frac{\gamma}{4}$, i.e., the ``exceptional point'' where $|\vec{h}_{\rm eff}| = \sqrt{(h_x)^2 + (-i \hbar \frac{\gamma}{4})^2} = 0$, and hence
\begin{align}
e^{-\frac{i}{\hbar} \hat{H}_{\rm eff} \tau} = e^{-\frac{\gamma}{4} \tau} \left\{ \mathbb{1}_2 - i \frac{ \gamma \tau}{4} (\hat{\sigma}_x-i \hat{\sigma}_z) \right\} .
\end{align}
The jumptime evolution (\ref{Eq:jumptime_evolution_equation}) now persists,
\begin{align}
\rho_{n+1}=\ket{0}\bra{0} ,
\end{align}
where the ground state becomes the steady state right after the first jump, with the waiting time distribution [cf.~(\ref{Eq:waiting_time_distribution})]
\begin{align}
w_{n \rightarrow n+1}(\tau) = \frac{\gamma^3 \tau^2}{16} e^{-\frac{\gamma}{2} \tau} \hspace{2mm} , \hspace{1mm} n \ge 1 .
\end{align}
Similar jumptime evolutions hold for general $h_x \neq 0$ (or any Hamiltonian that does not commute with $\ket{0}\bra{0}$). In contrast, the stationary states of the corresponding walltime master equation lie, for different Hamiltonians, on the surface of an ellipsoid in the Bloch sphere \cite{Sauer2013optimal}. Hence, the walltime evolution here lifts a degeneracy of the jumptime evolution in the Hamiltonian sector.

Alternatively, we can remove the dark state by adding the excitation process
\begin{align}
\hat{L}' = \sqrt{x} \hat{\sigma}_+ = \sqrt{x} \, \ket{1} \bra{0}
\end{align}
as a second jump operator, where $x >0$ denotes the ratio between the rates of the two processes. For $\hat{H} = 0$, we then obtain the effective Hamiltonian
\begin{align}
\hat{H}_{\rm eff} = -i \hbar \frac{\gamma}{4} \big( 1+x \big) \mathbb{1}_2 -i \hbar \frac{\gamma}{4} \big( 1-x \big) \hat{\sigma}_z ,
\end{align}
the conditioned time evolution operator
\begin{align}
e^{-\frac{i}{\hbar} \hat{H}_{\rm eff} \tau} = e^{-\frac{\gamma}{2} x \tau} \ket{0}\bra{0} + e^{-\frac{\gamma}{2} \tau} \ket{1}\bra{1} ,
\end{align}
and the jumptime evolution
\begin{align}
\rho_{n+1} = \hat{\sigma}_- \rho_n \hat{\sigma}_+ + \hat{\sigma}_+ \rho_n \hat{\sigma}_- .
\end{align}
The latter describes an ongoing population inversion for any initial state outside of the $x$-$y$ plane (which functions as a mirror plane) of the Bloch sphere. Note that the jumptime evolution here exhibits a degeneracy in the sector of the jump operators: it is independent of the rate ratio $x$. Moreover, in contrast to the stationary states in walltime, the evolution can here assume cyclic asymptotic behavior (which here takes the period $2$). At the same time, the waiting time distribution
\begin{align}
w_{n \rightarrow n+1}(\tau) = \gamma e^{-\gamma \tau} \bra{1} \rho_n \ket{1} + \gamma x e^{-\gamma x \tau} \bra{0} \rho_n \ket{0}
\end{align}
is $x$ dependent and state dependent, as well as the corresponding average waiting time $\overline{\tau}_{n \rightarrow n+1} = \int_0^\infty d \tau \, \tau \, w_{n \rightarrow n+1}(\tau)$,
\begin{align}
\overline{\tau}_{n \rightarrow n+1} = \frac{1}{\gamma} \bra{1} \rho_n \ket{1} + \frac{1}{\gamma x} \bra{0} \rho_n \ket{0} .
\end{align}
The resulting average duration of a cycle,
\begin{align}
T = \overline{\tau}_{n \rightarrow n+1} + \overline{\tau}_{n+1 \rightarrow n+2} = \frac{1}{\gamma} + \frac{1}{\gamma x} ,
\end{align}
is state-independent, and can be tuned arbitrarily large by diminishing the rate ratio $x$.

\subsection{Two-level system with dephasing}

Another paradigmatic process is the dephasing of a qubit, described by the single Lindblad operator
\begin{align}
\hat{L} = \hat{\sigma}_z .
\end{align}
Clearly, there are no dark states, irrespective of the Hamiltonian. For a general Hamiltonian $\hat{H}=\vec{h} \cdot \vec{\sigma}$, the effective Hamiltonian (\ref{Eq:effective_Hamiltonian}) reads 
\begin{align}
\hat{H}_{\rm eff} = -i \hbar \frac{\gamma}{2} \mathbb{1}_2 + \vec{h} \cdot \vec{\sigma} ,
\end{align}
and the conditioned time evolution operator is given by
\begin{align}
e^{-\frac{i}{\hbar} \hat{H}_{\rm eff} \tau} = e^{-\frac{\gamma}{2} \tau} \left\{ \cos \left[ \frac{\tau h}{\hbar} \right] \mathbb{1}_2 - i \frac{\tau}{\hbar} \sinc \left[ \frac{\tau h}{\hbar} \right] \vec{h} \cdot \vec{\sigma} \right\} ,
\end{align}
with $h = |\vec{h}|$ and $\sinc \, x \equiv x^{-1} \sin x$. The jumptime evolution (\ref{Eq:jumptime_evolution_equation}) can be determined analytically for arbitrary $\vec{h}$, and the waiting time distribution takes, for any $\vec{h}$, the state-independent form (\ref{Eq:state-independent_waiting_time}). For simplicity we focus on $\hat{H}= h_z \hat{\sigma}_z$, in which case we obtain
\begin{align} \label{Eq:dephasing_jumptime_evolution}
\rho_{n+1} =& \hat{\sigma}_z \rho_n \hat{\sigma}_z + \frac{2 h_z^2}{4 h_z^2+\hbar^2 \gamma^2} \big( \rho_n - \hat{\sigma}_z \rho_n \hat{\sigma}_z \big) \nonumber \\
& + \frac{\hbar \gamma h_z}{4 h_z^2+\hbar^2 \gamma^2} i[\hat{\sigma}_z, \rho_n] ,
\end{align}
i.e., the jumptime evolution depends both on $h_z$ and $\gamma$. Equation~(\ref{Eq:dephasing_jumptime_evolution}) can be brought into manifest Kraus form, but for the sake of compactness we choose this representation.

The walltime evolution equation which parallels (\ref{Eq:dephasing_jumptime_evolution}), $\partial_t \rho_t = -\frac{i}{\hbar} [h_z \hat{\sigma}_z, \rho_t] + \gamma \big( \hat{\sigma}_z \rho_t \hat{\sigma}_z - \rho_t \big)$, is solved by $\bra{1} \rho_t \ket{1} = 1 - \bra{0} \rho_t \ket{0} = \bra{1} \rho_0 \ket{1}$ and $\bra{1} \rho_t \ket{0} = \bra{1} \rho_0 \ket{0} \exp(-\frac{i}{\hbar} 2 h_z t - 2 \gamma t)$, describing states that spiral into the $z$ axis of the Bloch sphere; in particular, the state purity $r_t = \Tr[\rho_t^2] = \bra{0} \rho_0 \ket{0}^2 + \bra{1} \rho_0 \ket{1}^2 + 2 |\bra{0} \rho_0 \ket{1}|^2 \exp(-4 \gamma t)$ monotonically decreases for any initial state outside of the $z$ axis of the Bloch sphere, and irrespective of $h_z$.

To see how the jumptime evolution can deviate from this behavior, we evaluate (\ref{Eq:dephasing_jumptime_evolution}) for $\hat{H}=0$, which yields
\begin{align}
\rho_{n+1} = \hat{\sigma}_z \rho_n \hat{\sigma}_z .
\end{align}
This is solved by $\rho_n = \hat{\sigma}_z^n \rho_0 \hat{\sigma}_z^n$, which again describes a cyclic evolution with period $2$ for any initial state outside of the $z$ axis of the Bloch sphere. Remarkably, we find that the evolution is unitary and hence the respective state purity conserved, $r_n = \Tr[\rho_n^2] = \Tr[\rho_0^2]$, in stark contrast to the monotonous purity decay in walltime. This difference can be understood in terms of the parity of the number of quantum jumps, which is controlled under the jumptime evolution, while it is increasingly washed out under the walltime dynamics.

On the other hand, in the limit $h_z \gg \hbar \gamma$, the jumptime evolution (\ref{Eq:dephasing_jumptime_evolution}) can be approximated as
\begin{align}
\rho_{n+1} \approx \frac{1}{2} \rho_n + \frac{1}{2} \hat{\sigma}_z \rho_n \hat{\sigma}_z ,
\end{align}
which implies that the (initial state-dependent) minimum value of the purity $r_{\rm min} = \bra{0} \rho_0 \ket{0}^2 + \bra{1} \rho_0 \ket{1}^2$ is reached after a single jumptime step. Here, the rapid purity decay under the jumptime evolution can be traced back to the mixing of the completely randomized phases that are accumulated between initial state preparation and the first quantum jump (or any consecutive quantum jumps).

\subsection{Damped harmonic oscillator}

In this (infinite-dimensional state-space) example, the Hamiltonian is given by $\hat{H} = \hbar \omega (\hat{a}^{\dagger} \hat{a}+1/2)$, where the annihilation operator $\hat{a}$ also represents the single jump operator,
\begin{align}
\hat{L} = \hat{a} .
\end{align}
This describes, e.g., an oscillator in a zero temperature bath, or a lossy cavity mode. The effective Hamiltonian (\ref{Eq:effective_Hamiltonian}),
\begin{align}
\hat{H}_{\rm eff} = \hbar \omega \big( \hat{a}^{\dagger} \hat{a}+\frac{1}{2} \big) - i \hbar \frac{\gamma}{2} \hat{a}^{\dagger} \hat{a} ,
\end{align}
is then diagonal in the Fock basis $\ket{m}$, since $\hat{a}^{\dagger} \hat{a} \ket{m} = \hat{a}^{\dagger} \sqrt{m} \ket{m-1} = m \ket{m}$. Moreover, the ground state $\ket{0}$ describes a dark state, as $\hat{a} \ket{0}=0$ and $[\hat{H}, \ket{0}\bra{0}]=0$.

If we evaluate the jumptime evolution (\ref{Eq:jumptime_evolution_equation}) in the Fock basis, we obtain
\begin{subequations}
\begin{align}
\bra{m} \rho_{n+1} \ket{m'} = K(m,m') \bra{m+1} \rho_{n} \ket{m'+1} ,
\end{align}
with the propagator
\begin{align}
K(m,m') = \frac{2 \gamma \sqrt{(m+1) (m'+1)}}{(2+m+m') \gamma - 2 i \omega (m'-m)} .
\end{align}
\end{subequations}
The explicit solution then reads
\begin{align} \label{Eq:Damped_oscillator_solution}
\bra{m} \rho_{n} \ket{m'} = K(m,m')^n \bra{m+n} \rho_0 \ket{m'+n} ,
\end{align}
where $K(m,m)=1$.

Due to the presence of the dark state, the trace of the jumptime-evolved state is in general not preserved: $\Tr[\rho_{n}] = \sum_{m=0}^{\infty} \bra{m}\rho_n\ket{m} = \sum_{m=n}^{\infty} \bra{m}\rho_0\ket{m} \leq 1$. Instead, $\Tr[\rho_{n}]$ here describes the probability that $n$ jumps occur/can be observed, i.e., the fraction of quantum trajectories that arrive at the $n$th jump. Note that, similar to the two-level system under relaxation discussed above, the dark state is removed when adding the excitation process $\hat{L}' = \hat{a}^\dagger$.

If we evaluate (\ref{Eq:Damped_oscillator_solution}) for an initial Fock state $\rho_0 = \ket{N}\bra{N}$, we find that the jumptime evolution cascades down,
\begin{subequations} \label{Eq:Cascading_Fock_states}
\begin{align}
\rho_{n} = \ket{N-n}\bra{N-n} \hspace{2mm} {\rm for} \hspace{2mm} n \leq N ,
\end{align}
until it reaches the ground state (which is also the dark state of the system), where the jumptime evolution ends,
\begin{align}
\rho_{n}=0 \hspace{2mm} {\rm for} \hspace{2mm} n>N .
\end{align}
\end{subequations}
Let us stress again that the vanishing jumptime state implies that the system resides in the dark state. Moreover, we point out that the jumptime evolution preserves Fock states, which is not the case for the respective walltime evolution. As long as the evolution (\ref{Eq:Cascading_Fock_states}) is norm-preserving, $n \le N$, we can evaluate the waiting time distribution (\ref{Eq:waiting_time_distribution}), yielding
\begin{align}
w_{n \rightarrow n+1}(\tau) = \gamma (N-n) e^{-\gamma (N-n) \tau} \hspace{2mm} {\rm for} \hspace{2mm} n<N .
\end{align}
The corresponding average waiting times,
\begin{align}
\overline{\tau}_{n \rightarrow n+1} = \int_0^\infty d \tau \, \tau \, w_{n \rightarrow n+1}(\tau) = [\gamma (N-n)]^{-1} ,
\end{align}
grow as the ground state is approached.

\begin{figure}[htb]
\includegraphics[width=0.99\columnwidth]{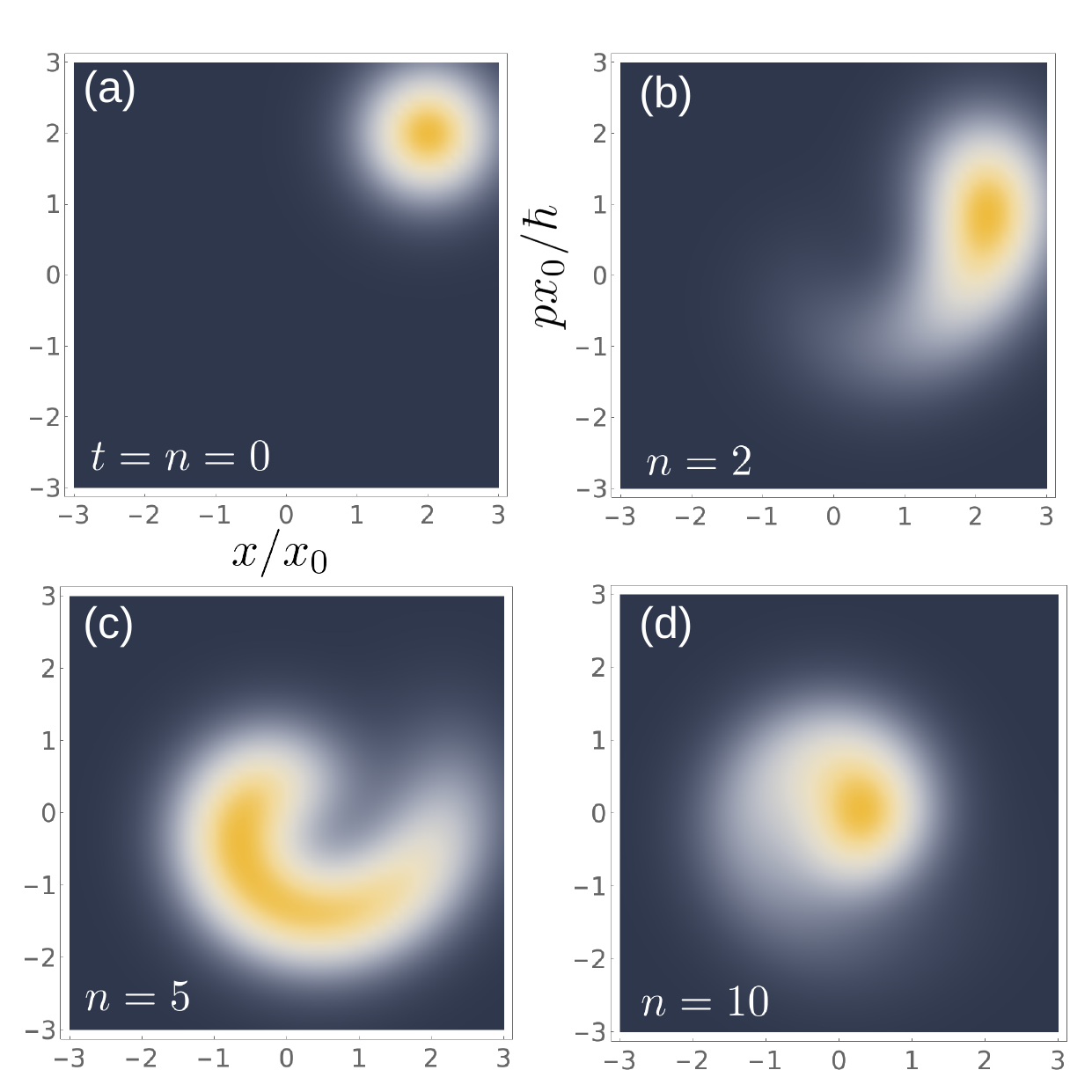}
\caption{\label{Fig:coherent_state} Jumptime evolution of a damped harmonic oscillator for an initial coherent state $\rho_0 = \ket{\alpha}\bra{\alpha}$, $\hat{a} \ket{\alpha} = \alpha \ket{\alpha}$, with $\alpha = 2 \, e^{i \pi/4}$. Shown are the Wigner functions of $\rho_n$, $W_n(x,p)$, for the jump counts $n=0$ (a), $n=2$ (b), $n=5$ (c), and $n=10$ (d). In contrast to the corresponding walltime evolution, the jumptime-evolved state deviates from a coherent state, until it arrives at the ground state. The latter is a dark state of the damped harmonic oscillator and hence acts as a sink on the trace of the state. The state norm is monotonously decreasing from (a) to (d), while the color code is gauged with respect to the maximum in each plot. The annihilation operator $\hat{a}$ and the phase space variables are connected through $\hat{a} = (\hat{x}/x_0 + i \hat{p} x_0/\hbar)/\sqrt{2}$, where $x_0=\sqrt{\hbar/\omega m}$ with mass $m$, and $\omega/\gamma=1$.}
\end{figure}

The jumptime evolution for an initial coherent state, $\rho_0 = \ket{\alpha} \bra{\alpha}$, where $\hat{a} \ket{\alpha} = \alpha \ket{\alpha}$, is depicted in Figure~\ref{Fig:coherent_state} in terms of the Wigner quasi-probability distribution \cite{Wigner1932quantum}. The Wigner functions of Gaussian states, which comprise coherent states, are known to be Gaussian-shaped phase space distributions. In our case, one easily sees that the jumptime-evolved states deviate from Gaussian distributions, implying that Gaussian states are not preserved under the jumptime evolution. In contrast, the corresponding walltime evolution, $\partial_t \rho_t = -\frac{i}{\hbar} [\hbar \omega \hat{a}^{\dagger} \hat{a}, \rho_t] + \gamma \big( \hat{a} \rho_t \hat{a}^\dagger - \frac{1}{2} \{ \hat{a}^\dagger \hat{a}, \rho_t \} \big)$, preserves Gaussian states, and coherent states evolve as $\alpha_t = \alpha_0 \exp(-i \omega t - \frac{\gamma}{2} t)$.

\subsection{Collisional decoherence}

As our final example, we consider a free particle, $\hat{H} = \hat{p}^2/2 m$ (with mass $m$), exposed to an environment exerting momentum kicks, e.g., a heavy test particle immersed in a background gas of light particles. It is assumed that the collisions merely decohere the particle state, without affecting its average momentum. We now have an infinite collection of jump operators, labeled by the momentum transfer $q$,
\begin{align}
\hat{L}_q = \sqrt{G(q)} \exp \left(i \frac{q \hat{x}}{\hbar} \right) ,
\end{align}
where the momentum transfer distribution $G(q) = G(-q) > 0$ determines the relative weight of the momentum kicks. We assume that $\int dq \, G(q) = 1$, which results in the effective Hamiltonian [cf.~(\ref{Eq:effective_Hamiltonian})]
\begin{align}
\hat{H}_{\rm eff} = \frac{\hat{p}^2}{2 m} - i \hbar \frac{\gamma}{2} \mathbb{1}_\infty .
\end{align}
There are no dark states present in this open continuous-variable system, implying an ongoing jump progression for any initial state. The state-independent waiting time distribution between jumps is given by (\ref{Eq:state-independent_waiting_time}).

Evaluating the jumptime evolution (\ref{Eq:jumptime_evolution_equation}) in momentum representation yields
\begin{subequations} \label{Eq:collisional_decoherence_jumptime}
\begin{align}
\bra{p} \rho_{n+1} \ket{p'} = \int dq \, G(q) K(p-q,p'-q) \bra{p-q} \rho_n \ket{p'-q} ,
\end{align}
with the propagator
\begin{align}
K(p,p') = \left[ 1 + i \frac{(p-p')(p+p')}{2 m \hbar \gamma} \right]^{-1} .
\end{align}
\end{subequations}
We can use (\ref{Eq:collisional_decoherence_jumptime}) to characterize the evolution behavior in jumptime. In line with the walltime evolution, the momentum expectation value is invariant,
\begin{align}
\langle \hat{p} \rangle_{n+1}=\langle \hat{p} \rangle_{n} ,
\end{align}
while the momentum variance stroboscopically grows,
\begin{align}
\langle (\Delta \hat{p})^2 \rangle_{n+1} = \langle (\Delta \hat{p})^2 \rangle_n + \Delta_G^2 ,
\end{align}
with $\Delta_G^2 = \int dq \, q^2 \, G(q)$. Notice that this broadening reflects the collision-induced decoherence.

The resulting transport behavior is indicated by the position expectation value, which evolves in steps,
\begin{align} \label{Eq:jumptime_transport}
\langle \hat{x} \rangle_{n+1} = \langle \hat{x} \rangle_n + \frac{\langle \hat{p} \rangle_n}{m \gamma} ,
\end{align}
controlled by the average velocity and the overall jump rate. Note that the step size $\frac{\langle \hat{p} \rangle_0}{m \gamma}$ decreases with increasing rate $\gamma$. With the average waiting time $\overline{\tau} = \int_0^\infty d \tau \, \tau \, w(\tau) = 1/\gamma$, we can rewrite (\ref{Eq:jumptime_transport}) as $\langle \hat{x} \rangle_{n+1} = \langle \hat{x} \rangle_n + \frac{\langle \hat{p} \rangle_0}{m} \overline{\tau}$, which parallels the transport according to the walltime evolution, $\langle \hat{x} \rangle_t = \langle \hat{x} \rangle_0 + \frac{\langle \hat{p} \rangle_0}{m} t$. However, while the latter cannot be distinguished from a free, isolated particle, the jumptime transport (\ref{Eq:jumptime_transport}) still reflects the presence of the collisions-inducing environment, by the resolution of the quantum jumps.

\section{Conclusions}

Based on the concept of quantum jump trajectories, we introduced jumptime-averaged quantum states, and demonstrated that these are governed by the discrete, deterministic jumptime evolution equation. The latter keeps track of the signature quantum jumps, in contrast to the deterministic, but jump-oblivious walltime master equation. While we put forward quantum trajectories as the connecting element between these two distinct ways of unfolding open quantum systems, both evolutions can be stated without reference to quantum trajectories, and each provides specific access to intrinsic properties of the open quantum system. In continuous measurement schemes, the two evolutions refer to different readout protocols, where the jumptime protocol is distinguished in that it actively involves the jump detection events.

Our examples show that the jumptime evolution can significantly deviate from its walltime counterpart. Notably, the jumptime evolution lifts generic degeneracies of the walltime evolution; on the other hand, the former can display robustness against perturbations in the Hamiltonian sector and/or the sector of the jump operators. Moreover, the jumptime evolution can exhibit cyclic asymptotic behavior, a characteristic trait that is not reflected by the corresponding Lindblad dynamics. Generally, jumptime unraveling may give new perspectives on, and insights into, for example, dissipative phase transitions, dissipative transport, non-Hermitian physics, quantum thermodynamics, and topological features in open systems (see, e.g., \cite{Gneiting2020unraveling}).

When realized with monitoring, the jumptime evolution promises to provide a versatile paradigm for engineering a broad class of quantum channels that can incorporate coherent and dissipative traits, with potential applications ranging from quantum sensing to quantum information processing (where ``gate switching'' may be triggered by jump detections, i.e., Hamiltonian and jump operators become conditioned on the jump count $n$, $\hat{H}^{(n)}$ and $\hat{L}_j^{(n)}$). Conceptually, jumptime averaging offers a new way to interpret and analyze continuous quantum measurements, be it in theory or experiment.

%\section*{Acknowledgments}
\paragraph*{Acknowledgments.}
F.N. was supported in part by:
Nippon Telegraph and Telephone Corporation (NTT) Research,
the Japan Science and Technology Agency (JST)
[via the Quantum Leap Flagship Program (Q-LEAP),
the Moonshot R\&D Grant No.~JPMJMS2061, and
the Centers of Research Excellence in Science and Technology (CREST) Grant No.~JPMJCR1676],
the Japan Society for the Promotion of Science (JSPS) [via the Grants-in-Aid for Scientific Research (KAKENHI) Grant~No.~JP20H00134 and the
JSPS–RFBR Grant~No.~JPJSBP120194828],
the Army Research Office (ARO) (Grant~No.~W911NF-18-1-0358),
the Asian Office of Aerospace Research and Development (AOARD) (via Grant~No.~FA2386-20-1-4069), and
the Foundational Questions Institute Fund (FQXi) via Grant~No.~FQXi-IAF19-06.
A.V.R. was partially supported by the Russian Foundation for Basic Research (RFBR Grant No.~19-02-00421 and RFBR-JSPS Grant No.~19-52-50015).

\bibliography{references}

\end{document}